\documentclass[review,12pt]{elsarticle}

\usepackage{amssymb}
\usepackage{natbib}
\usepackage{hyperref}

\journal{Journal of magnetism and Magnetic Materials}

\begin{document}

\begin{frontmatter}



\title{Single crystal study of layered U$_{n}$RhIn$_{3n+2}$ materials: case of the novel U$_{2}$RhIn$_{8}$ compound}


\author[mff]{Attila Bartha\corref{cor1}\fnref{fn1}}
\ead{bartha@mag.mff.cuni.cz}

\author[mff]{M. Kratochv\'{i}lov\'{a}}  
\author[fzu]{M. Du\v sek}
\author[mff]{M. Divi\v{s}}
\author[mff]{J. Custers}
\author[mff]{V. Sechovsk\'{y}}

\cortext[cor1]{Corresponding author}
\fntext[fn1]{Address: Charles University, Department of Condensed Matter Physics, Ke Karlovu 5, 121 16 Praha 2, Czech Republic; Tel. +420221911456}
\address[mff]{Department of Condensed Matter Physics, Charles University, Ke Karlovu 5, 121 16 Praha 2, Czech Republic}
\address[fzu]{Department of Structure Analysis, Institute of Physics ASCR, Cukrovarnick\'{a} 10, 162 00 Praha 6, Czech Republic}


\begin{abstract}
We report on the single crystal properties of the novel U$_{2}$RhIn$_{8}$ compound studied in the context of parent URhIn$_{5}$ and UIn$_{3}$ systems. The compounds were prepared by In self-flux method. U$_{2}$RhIn$_{8}$ adopts the Ho$_{2}$CoGa$_{8}$-type structure with lattice parameters \mbox{\textit{a} $= 4.6056(6)$ \AA} and \mbox{\textit{c} $= 11.9911(15)$ \AA}. The behavior of U$_{2}$RhIn$_{8}$ strongly resembles that of the related URhIn$_{5}$ and UIn$_{3}$ with respect to magnetization, specific heat and resistivity except for magnetocrystalline anisotropy developing with lowering dimensionality in the series UIn$_{3}$ vs. U$_{2}$RhIn$_{8}$ and URhIn$_{5}$. U$_{2}$RhIn$_{8}$ orders antiferromagnetically below \textit{T}$_{\textrm{\tiny{N}}}$ $= 117$ K and exhibits a slightly enhanced Sommerfeld coefficient $\gamma = 47$ mJ$\cdot$mol$^{-1}\cdot$K$^{-2}$. Magnetic field leaves the value of N\'{e}el temperature for both URhIn$_{5}$ and U$_{2}$RhIn$_{8}$ unaffected up to 9 T. On the other hand, \textit{T}$_{\textrm{\tiny{N}}}$ is increasing with applying hydrostatic pressure up to 3.2 GPa. 
The character of uranium 5\textit{f} electron states of U$_{2}$RhIn$_{8}$ was studied by first principles calculations based on the density functional theory. The overall phase diagram of U$_{2}$RhIn$_{8}$ is discussed in the context of magnetism in the related URh\textit{X}$_{5}$ and U\textit{X}$_{3}$ \mbox{(\textit{X} = In, Ga) compounds.}
\end{abstract}


\end{frontmatter}


\section{Introduction}
\label{sec:introduction}

Magnetism of uranium compounds is characterized by the large spatial extent of the 5\textit{f} wave functions which perceive their physical surroundings more intensively compared to the localized behavior of 4\textit{f} electrons. Typical example of that is the 5\textit{f}-ligand hybridization causing nonmagnetic behavior in several compounds characterized by the distance between the nearest U ions far larger than the Hill limit \cite{hill}. When considering the U\textit{X}$_{3}$ (\textit{X} = \textit{p}-metal) materials, the size of the \textit{p}-atom is a very important parameter. In the case of smaller \textit{X}-ions (Si, Ge) \cite{usi31}, the \textit{p}-wave function decays slower at the U-site, resulting in strong 5\textit{f}-\textit{p} hybridization which leads to lack of magnetic ordering (UGe$_{3}$, USi$_{3}$) \cite{uge3, usi31, usi3} while  larger \textit{X}-ions (In, Pb) cause the hybridization to be weaker resulting in magnetic ground state (UIn$_{3}$, UPb$_{3}$) \cite{uin3, upb3}.

The U$_{n}$\textit{TX}$_{3n+2}$ (\textit{n} = 1, 2; \textit{T} = transition metal; \mbox{\textit{X} =} In, Ga) \cite{ikeda, schonert, sechovsky, bartha, matsumoto} compounds adopt the layered Ho$_{n}$CoGa$_{3n+2}$-type structure which consists of \textit{n} U\textit{X}$_{3}$ layers alternating with a \textit{TX}$_{2}$ layer sequentially along the [001] direction in the tetragonal lattice. They are isostructural with the thoroughly investigated Ce$_{n}$\textit{TX}$_{3n+2}$ \cite{pfleiderer} compounds known for their outstanding physical properties such as the coexistence of unconventional superconductivity and magnetism or non-Fermi liquid behavior. These families of compounds provide unique opportunity to study the effect of dimensionality on physical properties due to their layered tetragonal structure. Adding a layer of \textit{TX}$_{2}$ pushes the character of the structural dimensionality from 3D to more 2D.

Since the U$_{2}$RhIn$_{8}$ compound has not been reported yet, we focused in this paper on the structure study followed by investigation of magnetic, transport and thermodynamic properties with respect to applied magnetic fields and hydrostatic pressure. In order to study the evolution of ground state properties on the structural dimensionality, we also prepared and investigated single crystals of URhIn$_{5}$ and UIn$_{3}$.

\section{Experimental}
\label{sec:experimental}

Single crystals of UIn$_{3}$, URhIn$_{5}$ and U$_{2}$RhIn$_{8}$ have been prepared using In self-flux method. High-quality elements U (purified by SSE \cite{sse}), Rh (3N5) and In (5N) were used. The starting composition of U:In = 1:10, U:Rh:In = 1:1:25 and U:Rh:In = 2:1:25 were placed in alumina crucibles in order to obtain UIn$_{3}$, URhIn$_{5}$ and U$_{2}$RhIn$_{8}$, respectively. The crucibles were further sealed in evacuated quartz tubes. The ampoules were then heated up to 950 $^\circ$C, kept at this temperature for 10 h to let the mixture homogenize properly and consequently cooled down to 600 $^\circ$C in 120 h. After decanting, plate-like single crystals of U$_{2}$RhIn$_{8}$ (URhIn$_{5}$) with typical dimensions of \mbox{$1\times 0.5\times 0.3$ \textrm{mm}$^{3}$} ($1\times 1\times 0.5$ \textrm{mm}$^{3}$) were obtained. In case of UIn$_{3}$, however, our grow attempts led to growth of single crystals of typical masses \mbox{$<$ 0.1 mg.} The single crystal of UIn$_{3}$ ($2\times 2\times 2$ \textrm{mm}$^{3}$) suitable for the bulk measurements was obtained as a by-product of the URhIn$_{5}$ synthesis.

Homogeneity and chemical composition of the single crystals were confirmed by scanning electron microscope (Tescan MIRA I LMH SEM) equipped with energy dispersive \textit{X}-ray analyzer (Bruker AXS). The crystal structures were determined by single crystal \textit{X}-ray diffraction using \textit{X}-ray diffractometer Gemini, equipped with an Mo lamp, graphite monochromator and an Mo-enhance collimator producing Mo K$_{\alpha}$ radiation, and a CCD detector Atlas. Absorption correction of the strongly absorbing samples ($\mu \sim$ 50 mm$^{-1}$) was done by combination of the numerical absorption correction based on the crystal shapes and empirical absorption correction based on spherical harmonic functions, using the software of the diffractometer CrysAlis PRO. The crystal structures were solved by SUPERFLIP \cite{superflip} and refined by software Jana2006 \cite{jana}.

The electrical resistivity measurements were done utilizing standard four-point method down to 2 K in a Physical Property Measurement System (PPMS). The specific heat measurements down to 400 mK were carried out using the He3 option. Magnetization measurements were performed in a superconducting quantum interference device (MPMS) from 2 to \mbox{300 K/400 K} and magnetic fields \mbox{up to 7 T.}

To investigate the effect of hydrostatic pressure on the transition temperature \textit{T}$_{\textrm{\tiny{N}}}$, we measured the temperature dependence of electrical resistivity using a double-layered (CuBe/NiCrAl) piston-cylinder type pressure cell with Daphne 7373 oil as the pressure-transmitting medium \cite{pressure, daphne}. Pressures up to 3.2 GPa were reached.

In order to acquire information about formation of magnetic moments in U$_{2}$RhIn$_{8}$, we applied the theoretical methods based on the density functional theory. The electronic structure and magnetic moments were calculated using the latest version of APW+lo WIEN2k code \cite{wien2k}. The 5\textit{f} electrons form the Bloch states with non-integer occupation number. The spin-orbit coupling was included using second-order variational step \cite{coupling}. Since we found the smaller value of the total magnetic moment than expected, we applied the LSDA+U method \cite{wien2k} and tuned the effective U to obtain the required total magnetic moment. The electronic structure calculations were performed at experimental equilibrium. The calculations were ferromagnetic for the sake of simplicity, since we have no information about the character of the antiferromagnetic ground state.

\section{Results and discussion}
\label{sec:results}

The obtained diffraction patterns revealed the Ho$_{2}$CoGa$_{8}$- (HoCoGa$_{5}$)-type structure (P4/mmm) for U$_{2}$RhIn$_{8}$ (URhIn$_{5}$). Table \ref{cryst} summarizes the lattice parameters, atomic coordinates and the equivalent isotropic displacement parameters \textit{U}$_{\textrm{\tiny{eq}}}$. The refinement parameters of the obtained data for U$_{2}$RhIn$_{8}$ equal \textit{R}$_{\textrm{\tiny{int}}} = 0.076$, \textit{R}[\textit{F}$^{2} > 3\sigma F^{2}$] = 0.035, the largest peak/hole in difference Fourier map $\Delta\rho_{\textrm{\tiny{max}}} = 5.84$ e\AA$^{-3}$/$\Delta\rho_{\textrm{\tiny{min}}} = - 4.04$ e\AA$^{-3}$. For URhIn$_{5}$: \textit{R}$_{\textrm{\tiny{int}}} = 0.041$, \textit{R}[\textit{F}$^{2} > 3\sigma F^{2}$] = 0.022, $\Delta\rho_{\textrm{\tiny{max}}} = 2.55$ e\AA$^{-3}$/\mbox{$\Delta\rho_{\textrm{\tiny{min}}} = - 2.01$ e\AA$^{-3}$.}

\begin{table*}[ht]
\caption{\label{cryst} Lattice parameters, fractional atomic coordinates and isotropic or equivalent isotropic displacement parameters for U$_{2}$RhIn$_{8}$ and URhIn$_{5}$.}
\vspace{3mm}
\centering
\begin{tabular}{c c c c c c}
\textbf{U$_{2}$RhIn$_{8}$} & Atom & \textit{x} & \textit{y} & \textit{z} & \textit{U}$_{\textrm{\tiny{iso}}}$*/\textit{U}$_{\textrm{\tiny{eq}}}$\\
\hline
\textit{a} = 4.6056(6) \AA & U & 0.5 & 0.5 & 0.30883(7) & 0.0059(3)\\
\textit{c} = 11.9911(15) \AA & Rh & 0.5 & -0.5 & 0 & 0.0078(6)\\
& In(1) & 0.5 & 0 & 0.5 & 0.0080(5)\\
& In(2) & 0.5 & 0 & 0.12263(11) & 0.0091(4)\\
& In(3) & 0 & 0 & 0.30916(14) & 0.0079(4)\\
\hline
\textbf{URhIn$_{5}$} &  &  &  &  & \\
\hline
\textit{a} = 4.6210(5) \AA & U & 0 & 0 & 0 & 0.00474(19)\\
\textit{c} = 7.4231(7) \AA & Rh & 1 & 0 & 0.5 & 0.0059(4)\\
& In(1) & 0.5 & 0 & 0.30179(11) & 0.0078(2)\\
& In(2) & 0.5 & 0.5 & 0 & 0.0076(3)\\
\hline
\end{tabular}
\end{table*}

\begin{figure}[ht]
\centering
 \includegraphics[width=0.9\textwidth, keepaspectratio=true]{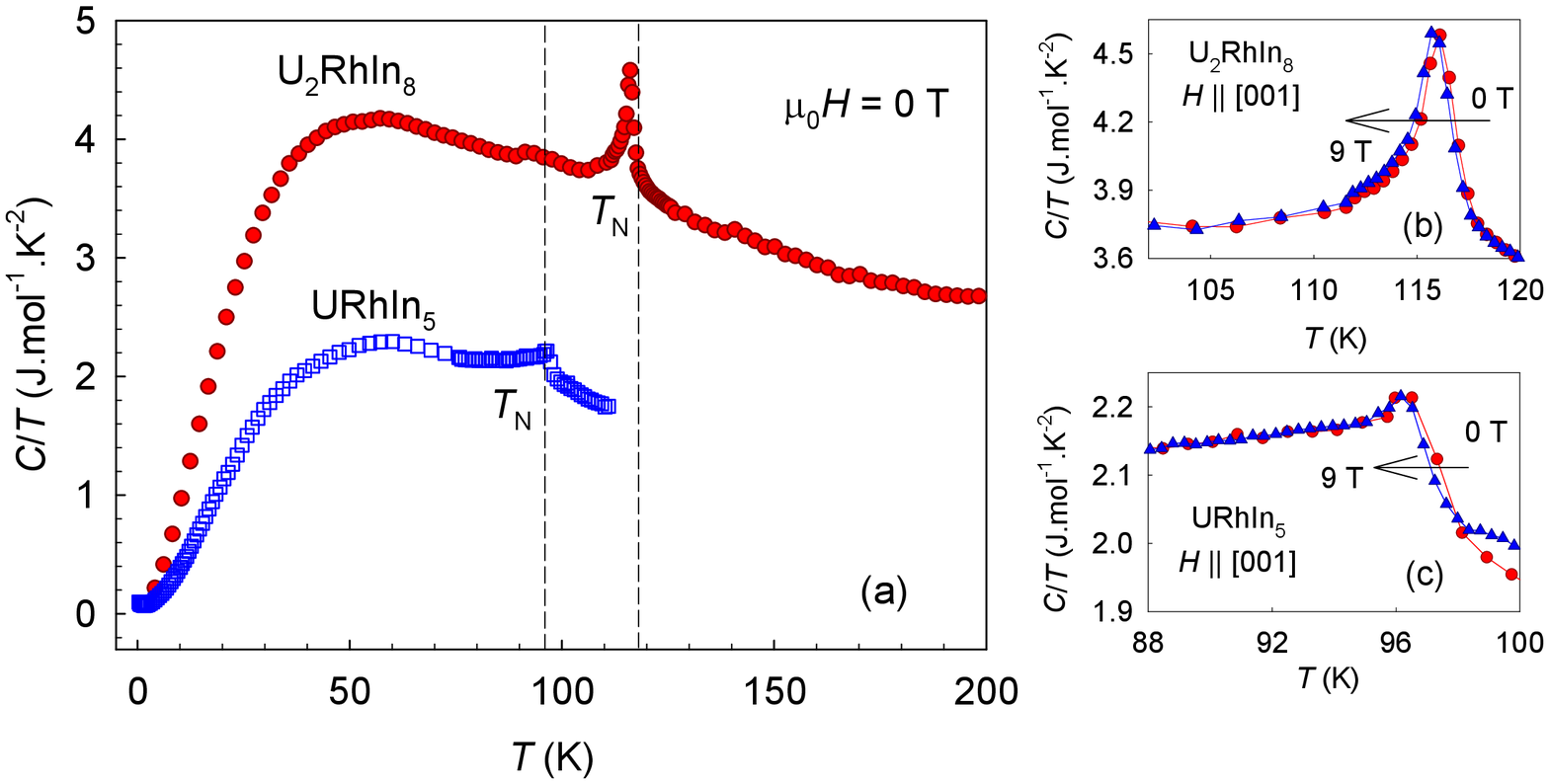}
 \caption{\label{fig:specific heat}Temperature dependence of the specific heat divided by temperature (a). The transition into the magnetically ordered state at \textit{T}$_{\textrm{\tiny{N}}}$ = 117 K (\textit{T}$_{\textrm{\tiny{N}}}$ = 98 K) for U$_{2}$RhIn$_{8}$ (URhIn$_{5}$) is marked by vertical dashed lines. Comparison of \textit{C}/\textit{T} for U$_{2}$RhIn$_{8}$ (b) and URhIn$_{5}$ (c) in zero and 9 T magnetic field, respectively, applied along the [001] axis.}
 \end{figure}
 
The temperature dependence of the specific heat \textit{C}(\textit{T}) divided by temperature for U$_{2}$RhIn$_{8}$ and URhIn$_{5}$ is presented in Fig. \ref{fig:specific heat}; a clear $\lambda$-shaped anomaly at \textit{T}$_{\textrm{\tiny{N}}}$ = 117 K and \textit{T}$_{\textrm{\tiny{N}}}$ = 98 K, respectively, indicates a second-order phase transition in both materials. Closer observation of the \textit{C}(\textit{T}) vs. \textit{T} curve of U$_{2}$RhIn$_{8}$ reveals a small anomaly at \mbox{\textit{T} $\sim$ 100 K}, which arises from a tiny amount of URhIn$_{5}$. The magnitude of the phonon contribution to the specific heat for both ternary compounds was determined from a $C/T = \gamma + \beta T^2$ fit to the data (fit interval 1 K $<$ \textit{T} $<$ 10 K). For the U$_{2}$RhIn$_{8}$ compound, the value of Sommerfeld coefficient yields \mbox{$\gamma$ = 47 mJ$\cdot$mol$^{-1}\cdot$U$\cdot$K$^{-2}$} and the $\beta$ coefficient equals to 3.4 mJ$\cdot$mol$^{-1}\cdot$K$^{-4}$ which corresponds to a Debye temperature \textit{T}$_{\textrm{\tiny{D}}}$ = 150 K. Sommerfeld coefficient of URhIn$_{5}$ equals to 60.7 mJ$\cdot$mol$^{-1}\cdot$U$\cdot$K$^{-2}$ while the $\beta$ coefficient yields the value of 3.3 mJ$\cdot$mol$^{-1}\cdot$K$^{-4}$ corresponding to the Debye temperature of 165 K. The values for URhIn$_{5}$ are close to those presented recently \cite{matsumoto}.
 
Fig. \ref{fig:specific heat} (b) represents data in applied magnetic field of 9 T  for U$_{2}$RhIn$_{8}$ along the [001] axis. The direct comparison with the zero field measurement reveals that \textit{T}$_{\textrm{\tiny{N}}}$ is almost unaffected within experimental uncertainty. Equivalent behavior is observed, Fig. \ref{fig:specific heat} (c), in the case of URhIn$_{5}$. Similar response to magnetic field is observed in the structurally related Ce-based compound, CeRhIn$_{5}$, where the transition temperature \textit{T}$_{\textrm{\tiny{N}}}$ tends to be rather insensitive to the application of magnetic field along the same direction \cite{cerhin5}.

\begin{figure}[!ht]
\centering
 \includegraphics[width=0.65\textwidth, keepaspectratio=true]{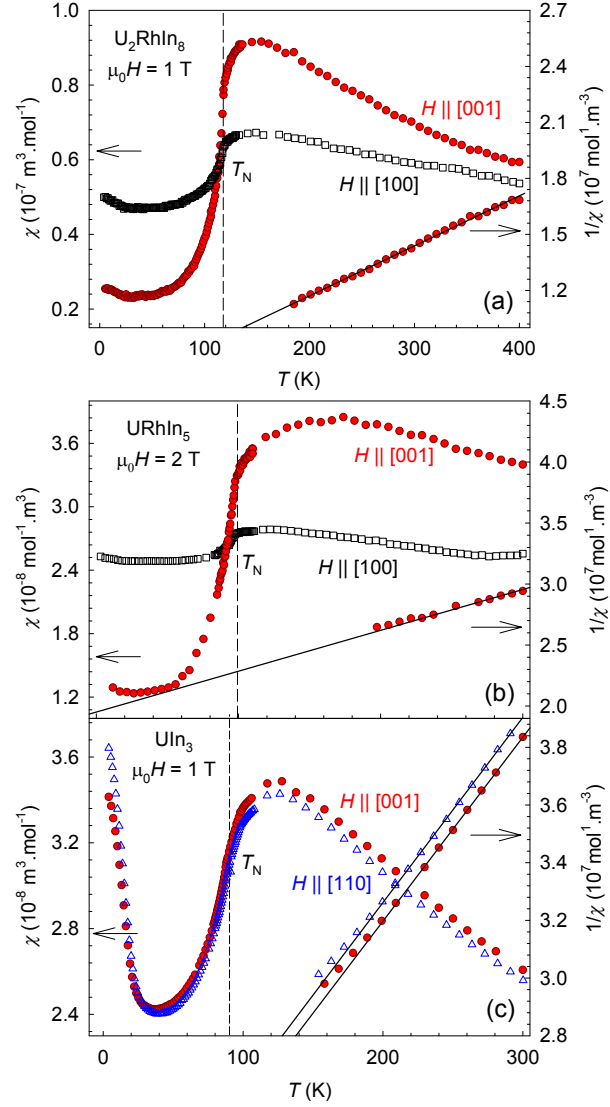}
 \caption{Temperature dependence of $\chi$(\textit{T}) and 1/$\chi$(\textit{T}) and Curie-Weiss fit of (a) U$_{2}$RhIn$_{8}$, (b) URhIn$_{5}$ and (c) UIn$_{3}$ for magnetic field oriented along the [100] and [001] direction in the case of U$_{2}$RhIn$_{8}$ and URhIn$_{5}$ along the [100] and [110] direction for UIn$_{3}$, respectively. Vertical dashed line marks the transition temperature \textit{T}$_{\textrm{\tiny{N}}}$.}
 \label{fig:magnetization}
\end{figure}

Fig. \ref{fig:magnetization} shows the temperature dependence of the $\chi$(\textit{T}) and 1/$\chi$(\textit{T}) of U$_{2}$RhIn$_{8}$ and URhIn$_{5}$ in magnetic field oriented along [100] and [001] directions and [110] and [001] in the case of UIn$_{3}$. The analysis of the $\chi$(\textit{T}) data lead to the determination of N\'{e}el temperatures as proposed by Fisher \cite{fisher}; the maximum of the $\partial(\chi(\textit{T}))/\partial\textit{T}$ curve. The behavior of the susceptibility curves resembles the one shown for URhIn$_{5}$ and UIn$_{3}$ \cite{bartha, uin3}, therefore we conclude that the phase transition drives the compound into an antiferromagnetic state. The magnetic susceptibility increases in all compounds with decreasing temperature and this increase is much pronounced for the [001] direction in the ternary compounds. The maximum value of susceptibility is reached at \textit{T}$_{\chi_{\textrm{\tiny{max}}}}$ = 130 K for UIn$_{3}$ (consistently with literature \cite{uin3}), at \mbox{\textit{T}$_{\chi_{\textrm{\tiny{max}}}}$ = 160 K} for URhIn$_{5}$ \cite{matsumoto} and at \textit{T}$_{\chi_{\textrm{\tiny{max}}}}$ = 150 K for U$_{2}$RhIn$_{8}$. Such behavior was previously observed in several different uranium compounds \cite{tateiwa} and it is generally supposed that this character of the susceptibility curves is associated with antiferromagnetic correlations when approaching \textit{T}$_{\textrm{\tiny{N}}}$. The value of \textit{T}$_{\chi_{\textrm{\tiny{max}}}}$ for both studied ternary compounds is the highest among uranium compounds up to our knowledge (UPd$_{2}$Al$_{3}$: \textit{T}$_{\chi_{\textrm{\tiny{max}}}}$ = 30 K; URu$_{2}$Si$_{2}$: \textit{T}$_{\chi_{\textrm{\tiny{max}}}}$ = 60 K \cite{galatanu}).

\begin{table}[!htb]
\caption{\label{tab:magnetic} N\'{e}el temperatures, effective magnetic moments obtained from Curie-Weiss fits and paramagnetic Curie temperatures for different orientation of magnetic fields for UIn$_{3}$ ([001] and [110] orientation), U$_{2}$RhIn$_{8}$ and URhIn$_{5}$.}
\vspace{3mm}
\centering
\begin{tabular}{c c c c}
\textbf{\textit{H} $\parallel$ [001]} & UIn$_{3}$ & U$_{2}$RhIn$_{8}$ & URhIn$_{5}$ \\
\hline
\textit{T}$_\textrm{\tiny{N}}$ (K) & 88 & 117 & 98\\
 $\mu_{\textrm{\tiny{eff}}}$ ($\mu_{\textrm{\tiny{B}}}$/U) & 3.16 & 3.45 & 3.6\\
 $\theta_{\textrm{\tiny{p}}} $(K) & $-300$ & $-240$ & $-400$\\
\hline
\textbf{\textit{H} $\parallel$ [110]} & & & \\
\hline
$\mu_\textrm{\tiny{eff}}$ ($\mu_{\textrm{\tiny{B}}}$/U) & 3.15 & - & -\\
$\theta_\textrm{\tiny{p}}$ (K) & $-310$ & - & -\\
\hline
\end{tabular}
\end{table}

In the vicinity of \textit{T}$_{\textrm{\tiny{N}}}$ = 117 K (98 K, 88 K) for U$_{2}$RhIn$_{8}$ (URhIn$_{5}$, UIn$_{3}$), a sharp drop of the magnetic susceptibility is observed in agreement with literature \cite{uin3, matsumoto, bartha}. This rapid decrease is again more pronounced in the [001] direction in the studied ternary compounds, pointing to the fact that the magnetic moments probably lie in this direction in the ordered state.

At low temperatures (\textit{T} $\sim$ 40 K) the susceptibility reaches its minimum value and rises up again. This increase is negligible in the case of URhIn$_{5}$ and U$_{2}$RhIn$_{8}$; however, it plays a dominant role in case of UIn$_{3}$. Previous studies of UIn$_{3}$ \cite{uin3} revealed similar behavior to that one presented for ternary compounds. This effect may have intrinsic nature or it could be also connected with a non-negligible amount of paramagnetic impurities, requiring further investigations. 

From the character of the 1/$\chi(T)$ curves above the N\'{e}el temperature it is evident, that the behavior of ternary compounds does not follow the Curie-Weiss law in \textit{H} $\parallel$ [100] direction due to magnetocrystalline anisotropy. The linear behavior appears probably above 400 K, as it is shown  i.e. for UPtGa$_{5}$ \cite{galatanu}, which we were able to confirm experimentally in the case of U$_{2}$RhIn$_{8}$ (see Fig. \ref{fig:magnetization} (a)).  
Such recovery of Curie-Weiss law at high temperatures indicates a localized nature of 5\textit{f} electrons. Thus, a crossover of the 5\textit{f} electrons from a low-temperature itinerant nature to a high-temperature localized one is observed. This crossover effect is characteristic for many heavy fermion compounds such as UPt$_{3}$, UPd$_{2}$Al$_{3}$ and URu$_{2}$Si$_{2}$ \cite{galatanu}.

In accordance with previous arguments, we applied the Curie-Weiss law in the \textit{H} $\parallel$ [001] direction for U$_{2}$RhIn$_{8}$ and URhIn$_{5}$, and in [001] and [110] direction in the case of UIn$_{3}$. We obtained qualitative values of the effective magnetic moments, summarized in table \ref{tab:magnetic}, that we were able to compare with those previously obtained \cite{matsumoto, bartha, uin3}.The large negative values of paramagnetic Curie temperatures reflect the huge uniaxial magnetocrystalline anisotropy in U$_{2}$RhIn$_{8}$ and URhIn$_{5}$ induced by anisotropic 5\textit{f}-ligand hybridization.

\begin{figure}[!htb]
\centering
 \includegraphics[width=0.75\textwidth, keepaspectratio=true]{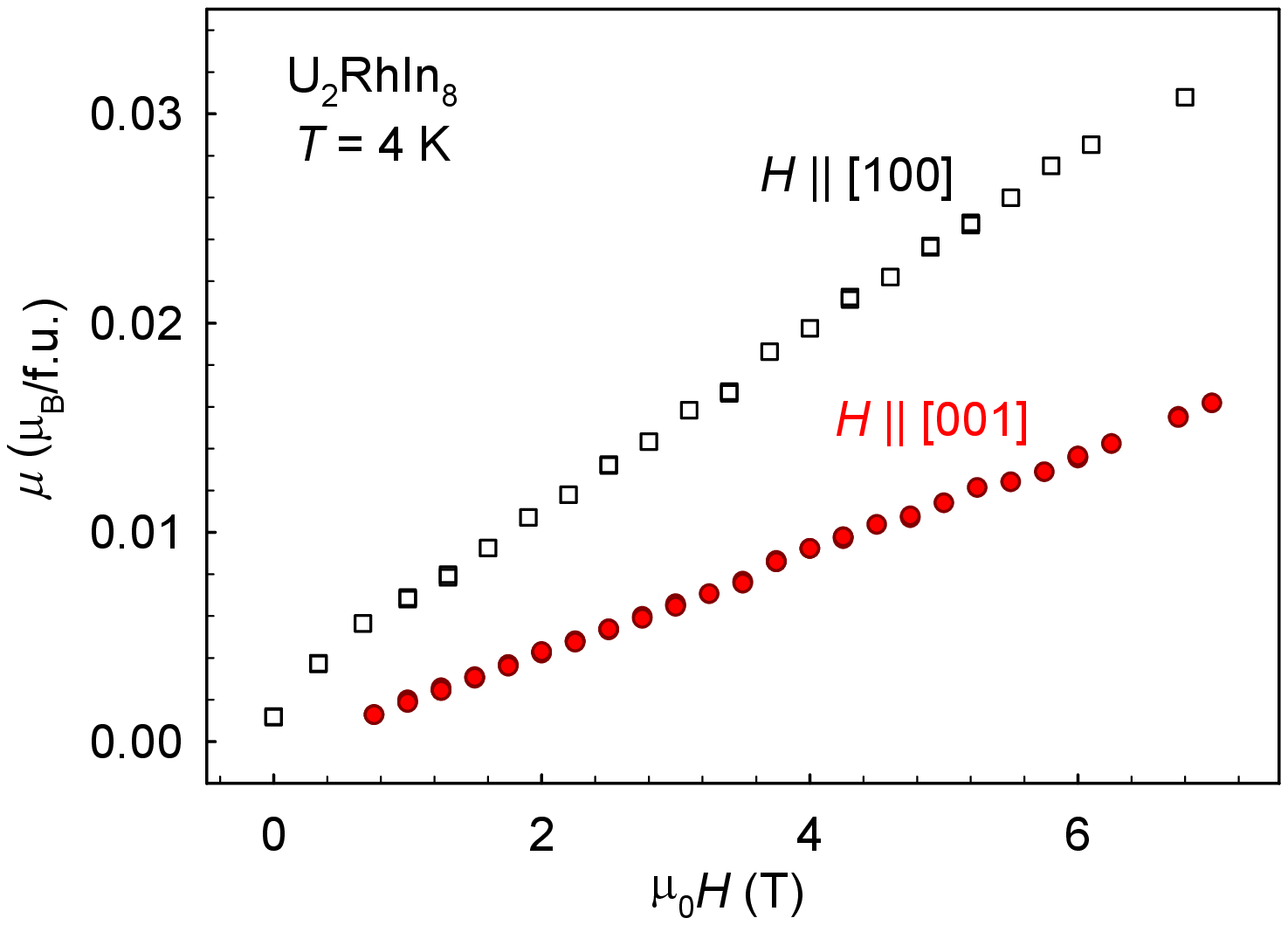}
 \caption{Magnetic field dependence of magnetization taken at 4 K of U$_{2}$RhIn$_{8}$ in magnetic field oriented along the [100] and [001] axes.}
 \label{fig:isothermal}
\end{figure}

The magnetic field dependence of magnetization (see Fig. \ref{fig:isothermal}) of U$_{2}$RhIn$_{8}$ was measured at \mbox{\textit{T} = 4 K} for magnetic field oriented along the [001] and [100] directions. Both magnetization curves reveal linear character up to 7 T; the [100] axis is almost twice higher than the magnetization in the other direction, which resembles the behavior of its more 2D counterpart \cite{matsumoto, bartha}. Using a relation \mbox{\textit{k}$_{\textrm{\tiny{B}}}$\textit{T}$_{\chi_{\textrm{\tiny{max}}}}$ $\simeq$ $\mu_{\textrm{\tiny{B}}}$\textit{H}$_{\textrm{\tiny{c}}}$} \cite{tateiwa}, where \textit{T}$_{\chi_{\textrm{\tiny{max}}}}$ = 150 K defines the position of the maximum of the magnetic susceptibility data, and \textit{H}$_{\textrm{\tiny{c}}}$ is the critical magnetic field of metamagnetic transition, we obtain a value of \mbox{\textit{H}$_{\textrm{\tiny{c}}}$ $=$ 220 T} for U$_{2}$RhIn$_{8}$. This extremely large value explains the absence of metamagnetic transition in our experimental data.

\begin{figure} [htp]
\centering
 \includegraphics[width=0.65\textwidth, keepaspectratio=true]{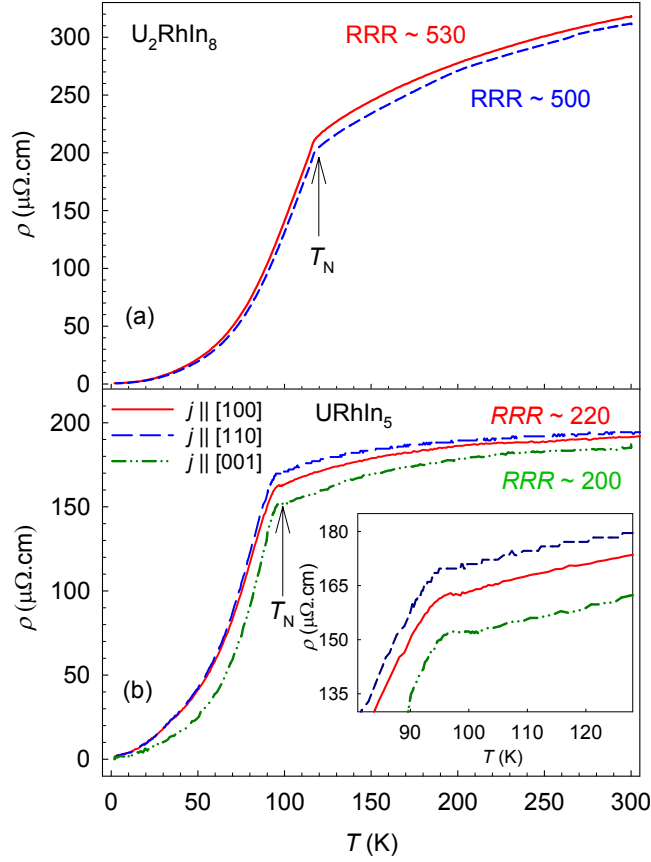}
 \caption{\label{fig:resistivity}Temperature dependence of the electrical resistivity for electrical current applied along [100] and [110] directions is shown for U$_{2}$RhIn$_{8}$ (a). The temperature dependence of the electrical resistivity for current applied along the [100], [110] and [001] directions for URhIn$_{5}$ compound (b). The inset in Fig. \ref{fig:resistivity} (b) shows the transition in detail. The arrows mark the onset of the antiferromagnetic transition at \textit{T}$_{\textrm{\tiny{N}}}$ = 117 K (\textit{T}$_{\textrm{\tiny{N}}}$ = 98 K) for U$_{2}$RhIn$_{8}$ (URhIn$_{5}$). One legend applies for both figures.} 
 \end{figure}

Fig. \ref{fig:resistivity} (a) shows the  temperature dependence of the electrical resistivity of U$_{2}$RhIn$_{8}$ for electrical current \textit{j} applied along the [100] and [110] axes. The room temperature resistivity equals 320 $\mu\Omega\cdot$cm along the [100] direction and is only slightly lower for the [110] direction \mbox{(310 $\mu\Omega\cdot$cm).} The residual resistivity ratio (RRR) exceeds 500 being a sign of a sample of very high quality. The electrical resistivity shows a monotonous decrease down to the value of the transition temperature \textit{T}$_{\textrm{\tiny{N}}}$. Near the transition temperature \mbox{\textit{T}$_{\textrm{\tiny{N}}}$ = 117 K}, a tiny kink is observed, accompanied by a second-order phase transition and a formation of a gap at the Fermi surface. Subsequently, the resistivity decreases rapidly with decreasing temperature. The low-temperature part of the electrical resistivity (2 K $<$ \textit{T}$_{\textrm{\tiny{fit}}}$ $<$ 30 K) can be fitted well using the equation appropriate for an energy gap ($\Delta$) antiferromagnet with an additional Fermi-liquid term, being: $\rho(T)=\rho_{0}+AT^{2}+DT(1+2T/\Delta)\textrm{exp}(-\Delta/T)$ \cite{resistivity}.
Optimal fitting of $\rho^{[100]}$ gives a residual resistivity value \mbox{$\rho_{0}$ = 0.56 $\mu\Omega\cdot$cm}, an electron-electron scattering coefficient of \textit{A} = 0.006 $\mu \Omega\cdot$cm$\cdot$K$^{-2}$, an electron-magnon and spin-disorder scattering prefactor \textit{D} = 1.1 $\mu \Omega\cdot$cm$\cdot$K$^{-1}$ and $\Delta$ = 118 K. Our fit yielded similar values of the parameters for current applied along the [110] direction.

Fig. \ref{fig:resistivity} (b) summarizes the overall temperature dependence of the electrical resistivity of URhIn$_{5}$ for electrical current \textit{j} applied along the [100], [110] and [001] axes. The room temperature resistivity equals 180 $\mu\Omega\cdot$cm in the basal plane and is only slightly lower for the [001]-axis direction \mbox{$\rho^{[001]}$ = 170 $\mu\Omega\cdot$cm}. Note that $\rho^{[100]}$ \mbox{(300 K)} is almost the same as reported by Matsumoto and co-workers \cite{matsumoto}. Similar to U$_{2}$RhIn$_{8}$, the RRR $\sim$ 200 of the URhIn$_{5}$ single crystals confirms their high quality. 
The data manifest distinct anomalies with onset at around \mbox{\textit{T} $\sim$ 100 K} for $\rho^{[001]}$ and at a slightly lower temperature \mbox{\textit{T} = 98 K} for $\rho^{[100]}$ and $\rho^{[110]}$, respectively, which is reminiscent of the N\'{e}el temperature anomaly for $\rho$(\textit{T}) in pure Cr \cite{chromium} - a spin-density-wave (SDW) antiferromagnet. Accordingly, the onset marks \textit{T}$_{\textrm{\tiny{N}}}$ and the increase in resistivity results from opening of the SDW gap. Below \textit{T}$_{\textrm{\tiny{N}}}$, the resistivity drops down rapidly in all directions. The low-temperature part was fitted according to similar formula as for U$_{2}$RhIn$_{8}$ with results \mbox{$\rho_{0}$ = 1 $\mu\Omega\cdot$cm}, \textit{A} = 0.013 $\mu \Omega\cdot$cm$\cdot$K$^{-2}$,  \textit{D} = 0.35 $\mu \Omega\cdot$cm$\cdot$K$^{-1}$ and \mbox{$\Delta$ = 82 K}. Our fit yielded similar values of the parameters for current along the [110] direction. However, in the case of \mbox{\textit{j} $\parallel$ [001]} we obtained a somewhat higher value of the gap energy \mbox{$\Delta$ = 119 K}.

After evaluating the data from specific heat and electrical resistivity measurements of U$_{2}$RhIn$_{8}$ and URhIn$_{5}$, the Kadowaki-Woods \cite{kadowaki} ratio \textit{A}/$\gamma^{2}$ could be calculated, where \textit{A} is the coefficient of the quadratic term in the temperature dependence of electrical resistivity. The value of 2$\cdot$10$^{-6}$ $\mu\Omega$cm$\cdot$(mol$\cdot$K/mJ)$^{2}$ for U$_{2}$RhIn$_{8}$ and 3.6$\cdot$10$^{-6}$ $\mu\Omega$cm$\cdot$(mol$\cdot$K/mJ)$^{2}$ for URhIn$_{5}$ was obtained, being one order of magnitude lower than in the common heavy-fermion systems \cite{kadowaki1}.

\begin{figure}[!htp]
\centering
\includegraphics[width=0.61\textwidth, keepaspectratio=true]{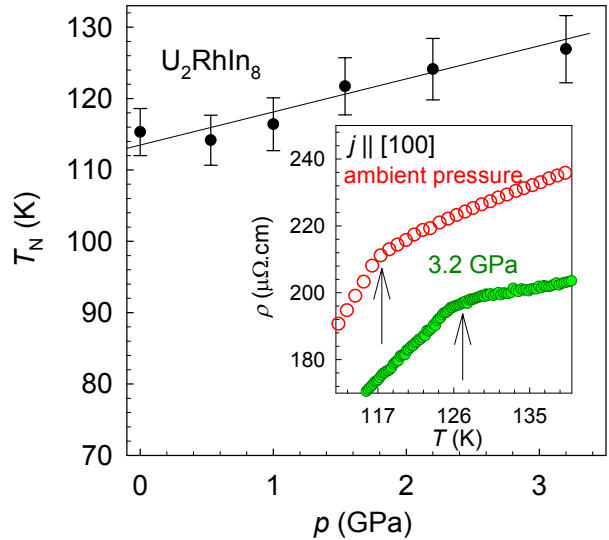}
 \caption{\label{fig:pressure} Temperature-pressure phase diagram of U$_{2}$RhIn$_{8}$. Inset shows the temperature dependence of the electrical resistivity for electrical current applied along the [100] direction at ambient pressure and at 3.2 GPa. Arrows indicate the transition at \textit{T}$_{\textrm{\tiny{N}}}$.}
\end{figure}

To investigate the effect of hydrostatic pressure, we performed electrical resistivity measurement of U$_{2}$RhIn$_{8}$ up to 3.2 GPa. The N\'{e}el temperature increases gradually with hydrostatic pressure, as is shown in the \mbox{\textit{T}-\textit{p}} phase diagram in Fig. \ref{fig:pressure}. A possible explanation of the positive pressure dependence of \textit{T}$_{\textrm{\tiny{N}}}$ is given by the spin-fluctuation theory of an itinerant 5\textit{f} electron system alongside with the Hubbard model \cite{cooper, sheng}. According to this scenario, hydrostatic pressure induces an increase in the hybridization between 5\textit{f} and conduction electrons, which strengthens the exchange coupling \textit{J} between U ions. On the other hand, it also decreases the 5\textit{f} magnetic moment at the uranium site. 
The value of \textit{T}$_{\textrm{\tiny{N}}}$ changes from 117 K to 128 K at ambient pressure and 3.2 GPa, respectively, with the rate of \mbox{5.4 $\pm$ 0.9 K$\cdot$GPa$^{-1}$.} This slope corresponds well to the pressure evolution of \textit{T}$_{\textrm{\tiny{N}}}$ in URhIn$_{5}$ \cite{matsumoto}. The detail of the resistivity curve near \textit{T}$_{\textrm{\tiny{N}}}$ for ambient pressure and for \mbox{3.2 GPa} with electrical current applied along the [001] axis is shown in the inset of Fig. \ref{fig:pressure}.

\begin{figure}[ht]
\centering
\includegraphics[width=0.65\textwidth, keepaspectratio=true]{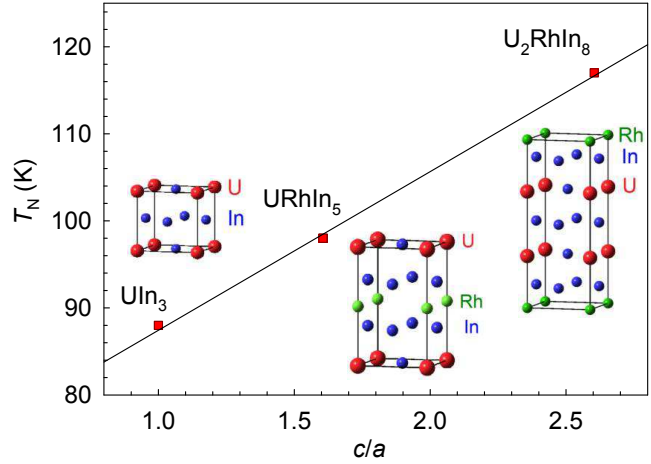}
 \caption{\label{fig:dimension} \textit{T}$_{\textrm{\tiny{N}}}$ vs. \textit{c}/\textit{a} diagram depicting the relation of the structural dimensionality and magnetism of uranium layered structures. The points are accompanied by the structures of UIn$_{3}$ (cubic), URhIn$_{5}$ (tetragonal) and U$_{2}$RhIn$_{8}$ (tetragonal) compounds. Solid black line is a guide to the eye.}
\end{figure}

The ternary compounds U$_{n}$\textit{TX}$_{3n+2}$ with \textit{T} $=$ Rh and \textit{X} $=$ In represent a system which offers the possibility to study the effect of structural dimensionality on the magnetism of uranium layered structures. Dimensionality develops from 3D towards 2D when spanning the series: UIn$_{3}$ (1-0-3) $\rightarrow$ U$_{2}$RhIn$_{8}$ (2-1-8) $\rightarrow$ URhIn$_{5}$ \mbox{(1-1-5).} In the other cases, either only the cubic compound (USn$_{3}$, UPb$_{3}$ \cite{usn3}) or the cubic compound (UGa$_{3}$ \cite{uga3}) together with its 1-1-5 (or 2-1-8) parent system (\textit{T} $=$ Fe, Co, Ni, Pd, Ir, Pt \cite{ufega5, urhga5, sechovsky, uptga5, u2rhga8}) are known. The Ga-based compounds with \textit{T} $=$ Fe, Rh, which form both 1-1-5 and \mbox{2-1-8} are paramagnetic. 
Since all the known U$_{n}$RhIn$_{3n+2}$ compounds order antiferromagnetically, a \textit{T}$_{\textrm{\tiny{N}}}$ vs. \textit{c}/\textit{a} diagram can be constructed (see Fig. \ref{fig:dimension}) in order to study the effect of structural dimensionality on the magnetic ordering. The evolution of transition temperature does not follow the dimensionality sequence "1-0-3 $\rightarrow$ 2-1-8 $\rightarrow$ \mbox{1-1-5"} as discussed by Cornelius \textit{et al.} \cite{cerhin5}. In contrast to the behavior in cerium compounds \cite{cerhin5}, the N\'{e}el temperatures of U$_{n}$RhIn$_{3n+2}$ reveal a monotonic evolution with respect to the \textit{c}/\textit{a} ratio. A possible explanation of this discrepancy is given by the different driving microscopic mechanisms in the compounds. The microscopic mechanisms in Ce-based compounds are mostly RKKY-type while in the uranium compounds the 5\textit{f}-ligand hybridization plays a substantial role. Moreover, the cerium compounds order magnetically well below 10 K while the ordering temperature in the case of U-based compounds is at least an order of magnitude higher.

The spin-polarized LSDA calculation splits the spin-up and spin-down bands with spin magnetic moment at uranium site 2.17 $\mu_{\textrm{\tiny{B}}}$. Since strong magnetocrystalline anisotropy is present, the spin-orbit coupling is also included into calculations. The calculated spin magnetic moment at uranium site decreases to \textit{M}$_{\textrm{\tiny{S}}}$ = 1.744 $\mu_{\textrm{\tiny{B}}}$ and the orbital magnetic moment \mbox{\textit{M}$_{\textrm{\tiny{L}}}$ = -2.418 $\mu_{\textrm{\tiny{B}}}$} is antiparallel. Magnetic moments located at rhodium and indium sites are both smaller than 0.1 $\mu_{\textrm{\tiny{B}}}$. Total uranium magnetic moment is \mbox{$\mid$\textit{M}$_{\textrm{\tiny{T}}}\mid$ = 0.674 $\mu_{\textrm{\tiny{B}}}$.} From comparison with URhIn$_{5}$ we expected larger total uranium moment about $\mid$\textit{M}$_{\textrm{\tiny{T}}}\mid$ = 1.6 $\mu_{\textrm{\tiny{B}}}$. Thus, the correlation movement of 5\textit{f} electrons is not negligible. Therefore we used LSDA+U method to describe correlated movement of 5\textit{f} electrons. Tuning effective Hubbard \textit{U} we have found spin magnetic moment \textit{M}$_{\textrm{\tiny{S}}}$ = 1.738 $\mu_{\textrm{\tiny{B}}}$ and orbital magnetic moment \textit{M}$_{\textrm{\tiny{L}}}$ = -3.3 $\mu_{\textrm{\tiny{B}}}$ providing the total magnetic moment \mbox{$\mid$\textit{M}$_{\textrm{\tiny{T}}}\mid$ = 1.592 $\mu_{\textrm{\tiny{B}}}$} for medium effective \mbox{\textit{U} = 1.3 eV.} We are fully aware that such calculation loses its first-principle character on this level, but on the other hand, we showed that these heuristically derived values of effective \textit{U} allow us to obtain valuable results.

\section{Conclusions}
\label{seq:conclusions}

Single crystals of UIn$_{3}$, URhIn$_{5}$ and the novel U$_{2}$RhIn$_{8}$ phase were synthesized using the In self-flux method and studied by means of magnetization, thermodynamic and transport measurements. The U$_{2}$RhIn$_{8}$ compound adopts the Ho$_{2}$CoGa$_{8}$-type structure with lattice parameters \mbox{\textit{a} = 4.6056(6) \AA} and \mbox{\textit{c} = 11.9911(15) \AA}. Measurements of specific heat and magnetization revealed a second-order phase transition into the antiferromagnetic state at \mbox{\textit{T}$_{\textrm{\tiny{N}}}$ = 117 K.} The compounds UIn$_{3}$ and URhIn$_{5}$ order antiferromagnetically below \textit{T}$_{\textrm{\tiny{N}}}$ = 88 K and 98 K, respectively, in accordance with literature \cite{uin3, matsumoto}. Electrical resistivity measurement revealed the very high quality of the studied ternary compounds with RRR exceeding 200 and 500, respectively. The temperature dependence of magnetic susceptibility for URhIn$_{5}$ and U$_{2}$RhIn$_{8}$ reveals strong magnetic anisotropy and suggests that both systems undergo an itinerant-localized crossover at high temperatures above 300 K, similar to other uranium-based compounds (UPd$_{2}$Al$_{3}$, URu$_{2}$Si$_{2}$ \cite{galatanu}), including UPtGa$_{5}$ \cite{galatanu} from the same group of compounds. The application of hydrostatic pressure supports the robustness of the antiferromagnetic phase in both ternary compounds with similar pressure coefficients \cite{matsumoto}. Successful synthesis of U$_{2}$RhIn$_{8}$ provides an opportunity among 5\textit{f} systems to study the evolution of ground state properties depending on \textit{c}/\textit{a} ratio. As is shown on Fig. \ref{fig:dimension} the evolution of \textit{T}$_{\textrm{\tiny{N}}}$ is monotonic with respect to \textit{c}/\textit{a} ratio. This behavior is in contrast with the cerium analogs \cite{cerhin5} and will be a subject of further investigation.

\section{Acknowledgement}
\label{seq:thanks}

Experiments were performed in MLTL (\url{http://mltl.eu/}), which is supported within the program of Czech Research Infrastructures (project no. LM2011025). Our work was supported by the Grant Agency of the Charles University (project no. 362214) and Czech Science Foundation (Project P203/12/1201). Single crystal \textit{X}-ray diffraction was performed in the Department of Structure Analysis, Institute of Physics ASCR.



\section{References}
\label{seq:references}

\end{document}